\documentclass[aps,prd,superscriptaddress,showpacs,twocolumn]{revtex4-1}
\usepackage{graphicx}
\usepackage[caption=false]{subfig}
\usepackage{epstopdf}
\usepackage{amsmath}
\usepackage{amsfonts} 
\usepackage{amssymb}
\usepackage{latexsym}
\usepackage{hyperref}
\usepackage[english]{babel}
\usepackage[utf8]{inputenc}
\usepackage[colorinlistoftodos]{todonotes}
\usepackage{xcolor}
\usepackage{slashed}
\usepackage{feynmp}
\usepackage{bm}
\usepackage{bbold}
\usepackage{eufrak}

\def\nn{\nonumber}
\def \pt{\partial}
\def \O{\mathcal{O}}

\def \M{\mathcal{M}}
\def \L{\mathcal{L}}

\begin{document}
\title{Spin- and velocity-dependent non-relativistic potentials in modified electrodynamics}

\author{G.P. de Brito}\email{gpbrito@cbpf.br}
\affiliation{Centro Brasileiro de Pesquisas F\'{i}sicas (CBPF), Rua Dr. Xavier Sigaud 150, Urca, Rio de Janeiro, Brazil, CEP 22290-180}

\author{P.C. Malta}\email{malta@thphys.uni-heidelberg.de}
\affiliation{Centro Brasileiro de Pesquisas F\'{i}sicas (CBPF), Rua Dr. Xavier Sigaud 150, Urca, Rio de Janeiro, Brazil, CEP 22290-180}
\affiliation{Institut f\"ur theoretische Physik, Universit\"at Heidelberg, Philosophenweg 16, 69120 Heidelberg, Germany}

\author{L.P.R. Ospedal}\email{leoopr@cbpf.br}
\affiliation{Centro Brasileiro de Pesquisas F\'{i}sicas (CBPF), Rua Dr. Xavier Sigaud 150, Urca, Rio de Janeiro, Brazil, CEP 22290-180}

\begin{abstract}
We investigate the interparticle potential between spin-0, -1/2 and -1 sources interacting in modified electrodynamics in the non-relativistic regime. By keeping terms of $\mathcal{O}( |{\bf p}|^2/m^2 )$ in the amplitudes, we obtain spin- and velocity-dependent interaction energies. We find well-known effects such as spin-orbit couplings, as well as spin-spin (dipole-dipole) interactions. For concreteness, we consider the cases of electrodynamics with higher derivatives (Podolsky-Lee-Wick) and hidden photons.
\end{abstract}

\maketitle

%%%%%%%%%%%%%%%%%%%%%%%%%%%%%%%

\section{Introduction} \label{sec1}

\indent

Classical Maxwell electrodynamics is a great theoretical and phenomenological success: its predictions are found to correctly describe a very wide variety of physical phenomena. In this scenario, the sources are agglomerations of electrons, protons and neutrons -- all spin-1/2 fermions -- so that the Maxwell equations actually encode the  macroscopic interaction between photons and spin-1/2 particles.

In the more general context of quantum field theories, however, the electromagnetic field is not bound to couple only to spin-1/2 sources, but may also interact with scalar (spin-0) and/or vector (spin-1) charged particles. At this point an interesting question arises: how can one distinguish between these sources via electromagnetic experiments? Or rather, can one find similarities -- or universalities -- among sources with different spins? A way to (partially) answer these questions is by investigating the potential energy between the spin-0, -1/2 and -1 sources.

The study of the interparticle potential serves yet another important cause, namely, the determination of the properties of the mediator, such as whether it is massive or not, as well as provide information on the number of propagating degrees of freedom. This is specially important in scenarios with physics beyond the Standard Model, whose low-energy effective theories generate modifications to usual electrodynamics.

The role of the sources, which couple to the gauge fields, is pivotal to understand -- and experimentally probe -- the features of modified electrodynamics. Here we wish to pursue a comparative study of the potential energy between scalar (spin-0), spinorial (spin-1/2) and vector (spin-1) sources, all electrically charged and massive. Similar analyses have been carried out e.g. in Refs.~\cite{Holstein1, Holstein2} -- also including higher-order and quantum effects -- in the context of pure Maxwell electrodynamics (and gravitation \cite{Holstein3}). Here we focus on a simpler approximation and work only with potential energies generated by one-boson exchange.

The potential energy being a classical, macroscopic quantity, it is natural that we work in the limit of small velocities, i.e., the non-relativistic (NR) limit. Most of the literature, e.g. Refs.~\cite{Nogueira0}-\cite{Nogueira}, work in the extreme NR limit of static sources. Though relevant, this restriction obscures the role of the spin of the sources, as can be easily seen in the case of spin-1/2 fermions, where the momentum is directly coupled to the spin matrices.

In order to exhibit the spin and momentum dependence in greater detail, we shall use the first Born approximation \cite{Maggiore}, where the potential energy is given by
\begin{equation}
E(r) = -  \int \frac{d^3 {\bf q}}{(2\pi)^3}  \, \mathcal{M}_{\rm NR} \,e^{i {\bf q}\cdot{\bf r}}, \label{born}
\end{equation}
with $\mathcal{M}_{\rm NR}$ being the NR limit of the fully relativistic Feynman amplitude $\mathcal{M}$.

\begin{figure}[h!]
\center
\includegraphics[width=7.5cm]{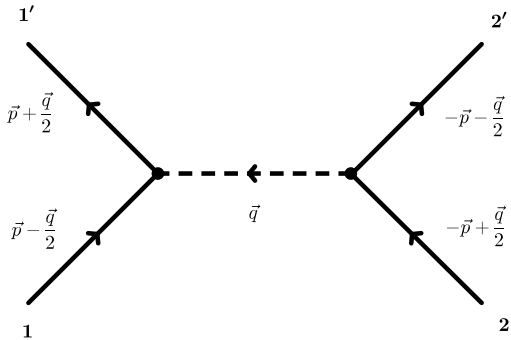}
\label{fig1}
\caption{Scattering process mediated by a vector boson (3-momenta displayed in the CM reference frame).} %\label{fig1}
\end{figure}

For the sake of concreteness, we shall work in the center-of-mass reference (CM) frame, so that only a pair of momentum variables is independent: $p = (p_1 + p'_1)/2 = (p_2 + p'_2)/2$ and $q = p'_1 - p_1 = -(p'_2 - p_2)$, thus representing the average momentum and the momentum transfer, respectively (cf. Fig.~\ref{fig1}). Moreover, we assume elastic scattering, where $q^0 = 0$, so ${\bf p}\cdot{\bf q} = 0$.

In order to bring to light spin (and momentum) dependence, it is necessary to go beyond $\mathcal{O}(1)$ in $|{\bf p}|/m$. At this level we have the well-known static $\sim e^{-Mr}/r$ spin-independent Yukawa potential. For this reason we shall work with amplitudes up to $\mathcal{O}(|{\bf p}|^2/m^2)$, where more interesting effects become evident. As we shall see, at this approximation level one obtains spin- and momentum-dependent generalizations of the Yukawa potential displaying, e.g. spin-orbit couplings as well as spin-spin interactions.

At this point it is important to briefly address the relationship between $\mathcal{M}_{\rm NR}$ and $\mathcal{M}$ in order to avoid confusion. Following Ref.~\cite{Maggiore}, we have
\begin{equation}
\mathcal{M}_{\rm NR} = \displaystyle \prod_{i = 1,2} \left(2E_i\right)^{-1/2}  \displaystyle \prod_{j = 1',2'} \left(2E_j\right)^{-1/2} \mathcal{M} \label{M_NR}
\end{equation}
and we note that the energy-dependent pre-factors are fundamental to obtain the NR potential correctly up to a given order. This is specially true here, as we want to go up to $\mathcal{O}(|{\bf p}|^2/m^2)$.

An important instance where this plays a role is in expressions involving combinations of energy and mass, such as $E + m$, which are momentum independent up to $\mathcal{O}(|{\bf p}|/m)$, but give $2m + |{\bf p}|^2/2m$ at the $\mathcal{O}(|{\bf p}|^2/m^2)$ level. In the case of spin-1/2 sources, for example, we shall see that such corrections affect the monopole-monopole terms, but not the spin-dependent ones  (cf. Section~\ref{sec_s1/2}).

%As mentioned above, we shall employ eqs.\eqref{born} and \eqref{M_NR} to compute the interaction potentials. Contrary to the standard textbook treatment up to $\mathcal{O}(|{\bf p}|/m)$, at the $\mathcal{O}(|{\bf p}|^2/m^2)$ level we must be more careful, specially with regard to the energy-dependent normalization of the wave functions in the transition from fully relativistic to the NR limit. In the case of spin-1/2 sources, for example, we shall see that such corrections affect the monopole-monopole terms, but not the spin-dependent ones. This issue will be further addressed in sec. \ref{sec_s1/2}.

%For the sake of concreteness we are going

We set out to study two phenomenologically interesting cases of modified electrodynamics. First we consider the Podolsky-Lee-Wick higher-order derivative model~\cite{Pod1}-\cite{LW2}, where a heavy ``ghost'' \`a la Pauli-Villars is introduced and helps tame UV divergences. Next we study the case of a massive neutral boson from a novel $U(1)_B$ symmetry kinetically mixed with the photon~\cite{Holdon}. Since usual matter is not charged under $U(1)_B$, this boson remains hidden, being dubbed a hidden (or dark) photon. The kinetic mixing induces a photon-hidden-photon oscillation and modifies the electromagnetic interaction between electrical charges~\cite{JJ}.

These two scenarios are representative and serve as applications for the more general discussion we wish to present. The main idea is to determine the role of the spin of the sources in the interparticle potential energy and, simultaneously, check for the signs of possible beyond the Standard Model effective models which may induce small corrections in the well-known Maxwell electrodynamics.

This paper is organized as follows: in Section~\ref{sec2}, we present a general and systematic discussion of the interparticle potentials for different sources, where we apply the NR approximation to the respective matter currents. In Section~\ref{sec3}, we analyze the standard Maxwell electrodynamics as a benchmark. In Sections~\ref{sec4} and \ref{sec5}, we work out the cases with the Podolsky-Lee-Wick modified electrodynamics and with photon-hidden-photon oscillations, respectively. We dedicate Section~\ref{sec6} to our concluding remarks. We use natural units ($c = \hbar = 1$) and a flat metric $\eta_{\mu\nu} = {\rm diag}(1,-1,-1,-1)$ throughout.

%%%%%%%%%%%%%%%%%%%%%%

\section{The non-relativistic limit of the interparticle potential} \label{sec2}

\indent

Let us set up our analysis by first considering the general structure of the NR potential energy for interactions mediated by neutral Abelian vector bosons. As discussed earlier, the interparticle potential energy is given by the first Born approximation, eq.\eqref{born}. This is essentially the NR limit of the quantum field theoretical scattering process between two particles (here labeled $1$ and $2$) with 3-momentum attributions as given in Fig.~\ref{fig1}.

Since the Feynman rules at tree level are equivalent to considering the matter currents $J^{\mu}$ associated with the interacting particles as being the interaction vertex, the amplitude may be written~\cite{epjc}
\begin{equation} \label{Feynman_amplitude_2}
\mathcal{M} = i J^{\mu}_{(1)}(p_1,p_1') \mathcal{P}_{\mu\nu}(q) J^{\nu}_{(2)}(p_2,p_2'), 
\end{equation}
where $J^{\mu}_{1(2)}$ are the currents associated with particles $1(2)$ and $\mathcal{P}_{\mu\nu}(k)$ is defined as the Feynman propagator, $D_{\mu\nu}(k)$, without its longitudinal part -- this is possible once we have $q_{\mu}J^{\mu}_{1(2)} = 0$, i.e., conserved currents \cite{Accioly1}. 

%Note that both currents generally depend on the incoming and outgoing momenta of the particles. 

%\begin{equation} \label{Feynman_amplitude_1}
%\mathcal{M} = i J^{\mu}_{(1)}(p,p') D_{\mu\nu}(k) J^{\nu}_{(2)}(q,q'), 
%\end{equation} where $D_{\mu\nu}(k)$ stands for the Feynman propagator

%This is based in the non-relativistic limit (i.e., $\textbf{p}/m_i \approx 0$) in the momentum conservation of the aforementioned process. 

%Since we are interested in the potential energy associated with conserved currents interacting via vector bosons, one can recast the Feynman amplitude as 
%\begin{equation} \label{Feynman_amplitude_2}
%\mathcal{M} = i J^{\mu}_{(1)}(p,p') \mathcal{P}_{\mu\nu}(k) J^{\nu}_{(2)}(q,q'), 
%\end{equation}
%where $\mathcal{P}_{\mu\nu}(k)$ is defined as the Feynman propagator without its longitudinal part \cite{Accioly1} -- this is possible once we have $k_{\mu}J^{\mu}_{1(2)} = 0$.

Before we treat particular cases for the external currents, we note that one can simplify the expression above, eq.\eqref{Feynman_amplitude_2}. If we consider that the field theory describing the vector boson which is responsible for the interaction at hand respects Lorentz invariance and receives no contribution from topological terms, we can express the Feynman propagator as a linear combination of the metric, $\eta_{\mu\nu}$, and momentum transfer, $q_\mu$.

Schematically one may then write
\begin{equation}
D_{\mu\nu}(q) = i a(q) \eta_{\mu\nu} + i b(q) q_\mu q_\nu, \label{prop_generic}
\end{equation}
where $a(q)$ and $b(q)$ are scalar functions specified by the particular theory under examination. When contracted with conserved currents, only the part linear in $\eta_{\mu\nu}$ survives and, as a consequence, one can write $\mathcal{P}_{\mu\nu}(q) = i a(q) \eta_{\mu\nu}$. Inserting this into eq.\eqref{Feynman_amplitude_2}, we finally get
\begin{equation}\label{Amplitude}
\mathcal{M} = - a(q) \eta_{\mu\nu} J^{\mu}_{(1)}(p_1,p_1') J^{\nu}_{(2)}(p_2,p_2').
\end{equation}

From now on we shall particularize the external currents to the cases of spin-0, -1/2 and -1 sources in order to derive suitable expressions to compute the potential energy as mediated via one-boson exchange.

%%%%%%%%%%%%%%%%%%%%%%%%%%%%%

\subsection{Spin-0 external currents} \label{sec_s0_cur}

\indent

We first consider the simplest case, where the charge carriers are described by spin-0 (complex) scalar fields. It is well known that the interaction vertex between an Abelian vector field and a scalar field is given by $V^\mu(p,p') = - i e (p'^\mu + p^\mu)$ with momenta flowing {\it into} the vertex. Here $e$ is the coupling constant.

Taking into account the usual Feynman rules and changing the final 4-momenta in favor of the momentum transfer and the initial 4-momenta, we obtain the following expression for the relativistic amplitude:
\begin{align}
\M &= - 4\,e_1 e_2\, a(q) \left(p_1 \cdot p_2 - \frac{1}{2}(p_1 - p_2)\cdot q - \frac{1}{4}q^2 \right) \nn\\
&= -4\,e_1 e_2 \, a(q) \left( E_1 E_2 + {\bf p}^2 \right),
\end{align}
where $E_1$ and $E_2$ stand for the energies of the incoming particles and we used that ${\bf p}_1 + {\bf p}_2 = 0$ in the CM frame, as well as ${\bf p}_{1(2)}^2 = {\bf p}^2 + {\bf q}^2/4$, since ${\bf p}\cdot{\bf q} = 0$. We note that we do not find extra $\sim {\bf q}^2$ terms.

According to eq.\eqref{M_NR}, the NR amplitude is 
\begin{eqnarray}
\M_{\rm NR} = -e_1 e_2 \, a(\textbf{q}) \left( 1  + \frac{\textbf{p}^2}{E_1 E_2} \right)
\end{eqnarray}
with the notation $a(\textbf{q}) = a(q)|_{q_0=0}$. Now, bearing in mind that
\begin{eqnarray}
E_{1(2)} \approx m_{1(2)} \left( 1 +  \frac{\textbf{p}_{1(2)}^2}{2m_{1(2)}^2}\right)
\end{eqnarray}
in the NR limit, we may recast $\M_{\rm NR}$ as
\begin{eqnarray} \label{Amp_scalar}
\M_{\rm NR} = -e_1 e_2 \, a(\textbf{q}) \left( 1  + \frac{\textbf{p}^2}{m_1 m_2} \right)  + \O\left(|\textbf{p}|^4/m^4\right).
\end{eqnarray}

Here it is important to stress that the energy-dependent pre-factors in eq.\eqref{M_NR} stem from normalization factors of the wave functions which compose the currents and, for this reason, we shall enforce our NR approximation by allowing terms up to and including $\mathcal{O}(|{\bf p}|^2/m^2)$ in $\mathcal{M}_{\rm NR} \sim \frac{J_1 J_2}{E_1 E_2}$. This means that terms such as $\sim {\bf p}^2{\bf q}^2/m^4$ will be disregarded as in eq.\eqref{Amp_scalar} above. This choice might lead us to not recover contact terms \cite{Holstein1}. Nevertheless, this approach is consistent with our approximation scheme and we shall make similar choices in the rest of this paper.

Finally, using eq.\eqref{born} along with eq.\eqref{Amp_scalar}, we obtain the following formal result for the NR potential energy associated with spin-0 charged particles
\begin{eqnarray}\label{energy_scalar}
E^{(\rm s = 0)}(r) =  e_1 e_2 \, \left( 1  + \frac{\textbf{p}^2}{m_1 m_2} \right) I_0(r),
\end{eqnarray}
with the integral $I_0(r)$ defined in Appendix~\ref{integrals}.

A similar analysis has been conducted in Ref.~\cite{Gupta} for the case of scalar electrodynamics. Our result, eq.\eqref{energy_scalar}, is different, as here we do not find a contact term (which is still different from the one found in Ref.~\cite{Holstein1}). The reason for this is simple: the two momentum variables we work with are ${\bf p}$ and ${\bf q}$, which are orthogonal. This ensures that they are independent from each other.

Since we are integrating over the momentum transfer, it is convenient to separate terms in the NR amplitude that depend on it from those which do not. This is most easily done with the orthogonal variables, ${\bf p}$ and ${\bf q}$. In Ref.~\cite{Gupta} the extra contact term arises as the initial momentum of one of the particles is used, and this momentum is not independent from the momentum transfer, which must be integrated according to eq.\eqref{born}.

\subsection{Spin-1/2 external currents} \label{sec_s1/2}

\indent

We now consider the case of external currents associated with charged spin-$1/2$ fields. The conserved vector current is 
\begin{equation}
J^{\mu}(p,p') = e \, \bar{u}(p') \gamma^\mu u(p) ,
\end{equation}
%and
%\begin{equation}
%J_{{(2)}}^{\mu}(q,q') = e_2 \bar{u}(q') \gamma^\mu u(q),
%\end{equation}
where $u(p)$ stands for the positive-energy solutions of the Dirac equation (its conjugate is $\bar{u} = u^{\dagger}\gamma^0$) and  $e$ is the coupling constant between the spin-1/2 fermion and the vector boson, as usual.

Using the standard Dirac representation for the gamma matrices, namely,
\begin{equation*}
\gamma^0 = \left( \begin{array}{cc}
\mathbf{1} &  \,\,\mathbf{0} \\
\mathbf{0} & - \mathbf{1}   \end{array} \right) \quad  \textmd{and} \quad \gamma^k = \left( \begin{array}{cc}
\,\,\mathbf{0} &  \sigma^k \\
- \sigma^k & \mathbf{0}   \end{array} \right),
\end{equation*}
one obtains
\begin{equation}
u( p_i) = \sqrt{E_i+m_i} \left( \begin{array}{c}
\xi \\\frac{\bm{\sigma}\cdot {\bf p}_i}{E_i+m_i} \xi  \end{array} \right), \label{spinor_basic}
\end{equation}
with $\xi$ being the basic spinor. We have normalized the spinor such that $ u^\dagger(p_i) u( p_i) = 2E_i$.

%$\xi = (1,0)^T$ or $\xi = (0,1)^T$, and $i=1,2$. 

%(we also have $\bar{u}( p) u( p) = 2m$).

Taking into account the momentum attributions from Fig.~\ref{fig1} and considering contributions up to second order in momenta, it follows that
\begin{equation}\label{current_1/2_0_1_1}
J^0_{(1)} = 2m_1e_1 \bigg[ 1 + \frac{1}{4m_1^2} \bigg( 2\textbf{p}^{\,2} + i (\textbf{q} \times \textbf{p}) \cdot \langle {\bm \sigma}_{(1)}\rangle \bigg) \bigg]
\end{equation}
and
\begin{equation}\label{current_1/2_i_1_1}
\textbf{J}_{(1)} = 2m_1e_1 \left( \frac{ \textbf{p} }{ m_1} - \frac{i}{2m_1} \textbf{q} \times \langle \bm{\sigma}_{(1)}\rangle \right),
\end{equation}
where we have used the notation $\langle \bm{\sigma}_{(i)}\rangle = \xi^\dagger \bm{\sigma}_{(i)} \xi$. Note that $J^\mu_{(2)}$ can be readily obtained from eqs.\eqref{current_1/2_0_1_1} and \eqref{current_1/2_i_1_1} by taking $e_1 \to e_2$, $m_1 \to m_2$, $\textbf{p} \to -\textbf{p}$, $\textbf{q} \to -\textbf{q}$ and $\langle \bm{\sigma}_{(1)}\rangle \to \langle \bm{\sigma}_{(2)}\rangle$.

%Using Eqs. (\ref{current_1/2_0_1_1}) and (\ref{current_1/2_i_1_1}), after some algebra we get
%\begin{eqnarray}
%	& &\eta_{\mu\nu}J^{\mu}_{(1)} J^{\nu}_{(2)} = e_1 e_2 \bigg\{ 1 + \frac{\textbf{p}^{\,2}}{4 m_1^2} + \frac{\textbf{q}^{\,2}}{4 m_2^2} - \frac{\textbf{p}\cdot \textbf{q}}{4 m_1 m_2} +\nn\\ &-& \frac{1}{2} \bigg( \frac{1}{2m_1} + \frac{1}{2m_2} \bigg)^2 \textbf{k}^{\,2} + i \textbf{k} \cdot \bigg( \frac{\textbf{p} \times \langle \textbf{\sigma}_{(1)}\rangle}{4m_1^2} - \frac{\textbf{q} \times \langle \textbf{\sigma}_{(2)}\rangle}{4m_2^2}\bigg) + \nn\\&+& i \textbf{k}\cdot \bigg( \frac{\textbf{p} \times \langle \textbf{\sigma}_{(2) }\rangle - \textbf{q} \times \langle \textbf{\sigma}_{(1) }\rangle}{2m_1 m_2}\bigg) + \nn \\&+& \frac{1}{4 m_1 m_2} \bigg( (\textbf{k}\cdot \langle \textbf{\sigma}_{(1) }\rangle ) (\textbf{k}\cdot \langle \textbf{\sigma}_{(2) }\rangle ) - \textbf{k}^{\,2} \langle \textbf{\sigma}_{(1)}\rangle \cdot \langle \textbf{\sigma}_{(2)}\rangle \bigg) \bigg\}.
%\end{eqnarray}

Spin-1/2 currents present a richer phenomenology -- as compared with spin-0 ones -- due to the presence of a non-trivial spin. This can be seen by plugging eqs.\eqref{current_1/2_0_1_1} and \eqref{current_1/2_i_1_1} in eq.\eqref{Amplitude}, which, in view of eq.\eqref{M_NR}, gives
\begin{equation}
\mathcal{M}_{\rm NR} = \mathcal{M}^{(0)}_{\rm NR} + \mathcal{M}^{(\rm vel)}_{\rm NR} + \mathcal{M}^{(\rm s-vel)}_{\rm NR} + \mathcal{M}^{(\rm s-s)}_{\rm NR},
\end{equation}
where
\begin{eqnarray*}
\mathcal{M}^{(0)}_{\rm NR} & = & -\,e_1 e_2\, a(q) \\
\mathcal{M}^{\rm (vel)}_{\rm NR} & = & - e_1 e_2 a(q) \left[ \frac{\textbf{p}^{\,2}}{m_1 m_2} - \frac{\textbf{q}^{\,2}}{8} \left( \frac{1}{m_1^2} + \frac{1}{m_2^2} \right) \right] \nonumber \\
\mathcal{M}^{\rm (s-vel)}_{\rm NR} & \!=\! &   -i e_1 e_2 \, a(q) \, \textbf{q} \! \cdot \! \bigg\{ \textbf{p} \times \left[ \frac{1}{4} \left( 
\frac{\langle \bm{\sigma}_{(1)}\rangle}{m_1^2} \!+\!
\frac{\langle \bm{\sigma}_{(2)}\rangle}{m_2^2} \right) \right. \!\!+\nonumber  \\ 
&&  \,\,\,\,\,\qquad \quad \,\,\,\,\,   + \left.
\frac{1}{2m_1 m_2} \, \left( \langle \bm{\sigma}_{(1) }\rangle + 
\langle \bm{\sigma}_{(2) }\rangle\right) \! \right] \bigg\} \\
\mathcal{M}^{\rm (s-s)}_{\rm NR} & = & -\frac{e_1 e_2 \, a(q)}{4 m_1 m_2}\bigg\{(\textbf{q}\cdot \langle \bm{\sigma}_{(1) }\rangle )(\textbf{q}\cdot\langle \bm{\sigma}_{(2) }\rangle ) + \nonumber \\
&&   \,\,\,\,\,\qquad \qquad\qquad \,\,\,\,\,\,     -    \textbf{q}^{\,2}\langle \bm{\sigma}_{(1)}\rangle\cdot\langle \bm{\sigma}_{(2)}\rangle \bigg\},
\end{eqnarray*}

Each of the above equations stands for a contribution of a different nature: $\mathcal{M}^{(0)}_{\rm NR}$ is the Feynman amplitude associated with static particles with no spin structure; $\mathcal{M}^{\rm (vel)}_{\rm NR}$ is the contribution dependent on the velocity of the charge carriers; $\mathcal{M}^{\rm (s-vel)}_{\rm NR}$ is related with spin-velocity (or spin-orbit) interactions and $\mathcal{M}^{\rm (s-s)}_{\rm NR}$ displays spin-spin interactions.

Finally, the NR potential energy is found to be
\begin{align}\label{Energia_fermion_2}
&E^{(\rm s = 1/2)}(r)  =  e_1 e_2 \,\bigg\{ \left( 1 + \frac{ \textbf{p}^2}{m_1 m_2}  \right) I_0(r) +\nonumber \\ 
&\quad -  \left[\frac{1}{8} \left( \frac{1}{m_1^2} + \frac{1}{m_2^2} \right) + \frac{1}{m_1m_2} \langle {\bf S}_{1} \rangle \cdot \langle {\bf S}_{2}\rangle  \right] I_1(r) +\nonumber \\
&\quad +  \frac{1}{r} I'_0(r) \,\textbf{L} \! \cdot\! \left( \frac{\langle {\bf S}_{1} \rangle}{2m_1^2} + \frac{\langle {\bf S}_{2} \rangle}{2m_2^2} + \frac{\langle {\bf S}_{1} \rangle + \langle {\bf S}_{2} \rangle}{m_1 m_2}\right) +\nonumber \\
&\quad +  \frac{1}{m_1m_2} \sum_{i,j=1}^3\langle {{\bf S}}_{1} \rangle_i \langle {{\bf S}}_{2}\rangle_j I_{ij}(r) \bigg\},
\end{align}
where we made ${\bf S} = {\bm \sigma}/2$ and ${\bf L} = {\bf r}\times{\bf p}$ is the (orbital) angular momentum. The integrals $I_0(r)$, $I_1(r)$ and $I_{ij}(r)$ are listed in Appendix \ref{integrals}.

%%%%%%%%%%%%%%%%%%%%%%%%%%%%%%%%%

\subsection{Spin-1 external currents} \label{sec_2c}

\indent

Lastly, we consider the less usual -- but nonetheless relevant -- case where the charge carriers are described by a (complex) spin-1 vector field, here designated by $W^{\mu}$. The conserved current is given by %\cite{helayel2}
\begin{eqnarray}
J^{\mu} &=& ie (W^{\mu\nu *} W_\nu - W^{\mu\nu} W^*_\nu) + \nonumber \\
&+& ie(g-1)\partial_{\nu} (W^\mu W^{\nu*}-W^{\mu*}W^\nu), \label{cur_W}
\end{eqnarray} 
where $W_{\mu\nu} = \partial_\mu W_\nu - \partial_\nu W_\mu$ is the Abelian field strength.

The first line of eq.\eqref{cur_W} stands for the usual Noether current coming from the global $U(1)$ symmetry. The second line plays an important role in the electrodynamics of spin-1 fields: it stems from a non-minimal coupling between the gauge field $A_\mu$ and the vector field $W_\mu$, namely, $\mathcal{L}_{\rm int} \supset ie(g-1)W^*_\mu W_\nu F^{\mu\nu}$, which is reminiscent from a broken $SU(2)\otimes U(1)$ symmetry. As argued in Refs.~\cite{Jackiw,Ferrara}, the addition of this term is mandatory in order to ensure the correct tree-level gyromagnetic factor for the vector field, i.e., $g=2$ (instead of $g=1$).

%Since in this paper we deal with higher derivative theories, we may wonder if is possible to include higher derivative terms in the non-minimal coupling. However, is not difficult to verify that higher derivative terms in the non-minimal coupling does not lead to any new contribution in the non-relativistic limit of the conserved current, which is the situation that we are concerned. The reason for this is simple: higher derivatives in the non-minimal coupling would lead to terms proportional (at least) to the third power in the momentum transfer and, since in the non-relativistic limit we consider only terms up to the second power in the momentum, any contribution from higher derivatives in the non-minimal coupling is irrelevant for the non-relativistic limit of the current.

The free vector field solution is %\cite{helayel2}
\begin{equation}
W^\mu(p_i) = N_W^{(1)} \left[\frac{\textbf{p}_i \cdot \bm{\epsilon}_i}{m_i},\, \bm{\epsilon}_i + \frac{1}{m_i (E_i +m_i)} (\textbf{p}_i\cdot\bm{\epsilon}_i)\,\textbf{p}_i \right], \label{W}
\end{equation}
where $\bm{\epsilon}_i$ stands for the polarization 3-vector and $|N_W^{(i)}|^2 = 1$, with $i=1,2$ labeling the sources (no summation implied). Equation \eqref{cur_W} may be rewritten in momentum space (for particle 1)
\begin{align}
& J_{(1)}^{\mu}(p,p') = e_1(p+p')^{\mu} W^\nu(p)W_\nu^*(p') +   \\
&\quad- e_1 g \left[p_\nu W^{\nu*}(p') W^\mu(p) + p'_\nu W^{\nu}(p) W^{\mu*}(p') \right]. \nonumber
\end{align}

%and
%\begin{equation}
%W^{\mu*}(p') = N_W^{(1)\ast} \bigg(\frac{\textbf{p}{\,'} \cdot \bm{\epsilon}^{\,*}}{m_1},\, \bm{\epsilon}^{\,*} + \frac{1}{2m_1^2} (\textbf{p}{\,'}\cdot\bm{\epsilon}^{\,*})\,\textbf{p}{\,'} \bigg),
%\end{equation}

%\begin{eqnarray}
%\mathcal{M}_{N\!R} &=& - e_1 e_2\, a(\textbf{k}) \bigg\{ 1 + \frac{1}{2}\bigg( \frac{\textbf{p}}{m_1} - \frac{\textbf{q}}{m_2} \bigg)^2 + i \textbf{k} \cdot \bigg[ \frac{g-1}{2m_2^2} \left( \textbf{q} \times \Sb \right) -  \frac{g-1}{2m_1^2} \left( \textbf{p} \times \Sa \right) + \nn \\ &+& \frac{g}{2m_1 m_2} \left( \textbf{q} \times \Sa - \textbf{p} \times \Sb \right) \bigg]  - \frac{g-1}{2} \left( \frac{1}{m_1^2} (\textbf{k}\cdot\textbf{\epsilon}_{(1)})(\textbf{k}\cdot\textbf{\epsilon}_{(1)}^{\,*}) + \frac{1}{m_2^2} (\textbf{k}\cdot\textbf{\epsilon}_{(2)})(\textbf{k}\cdot\textbf{\epsilon}_{(2)}^{\,*})\right) + \nn \\&+& \frac{g^2}{4m_1 m_2} \left( (\textbf{k}\cdot \Sa) (\textbf{k}\cdot \Sb) - \textbf{k}^{\,2} \Sa \cdot \Sb \right) \bigg\},
%\end{eqnarray}

In order to exhibit the spin dependence, we define the spin matrix, $(S_i)_{jk} = -i \varepsilon_{ijk}$, where $S_i$ is related to the vector representation of the rotation group, $\Sigma_{ij} = \varepsilon_{ijk} S_k$. Using eq.\eqref{W} and repeating the steps of the previous sections we find the NR potential energy,
\begin{align}\label{Energia_boson_2}
&E^{(\rm s = 1)}(r) = e_1 e_2 \,\bigg\{ \left( 1 + \frac{ \textbf{p}^2}{m_1 m_2}  \right) I_0(r) +\nonumber \\ 
&\quad -  \frac{1}{m_1m_2} \langle {\bf S}_{1} \rangle \cdot \langle {\bf S}_{2}\rangle I_1(r) +\nonumber \\
&\quad +  \frac{1}{r} I'_0(r) \,\textbf{L} \! \cdot\! \left( \frac{\langle {\bf S}_{1} \rangle}{2m_1^2} + \frac{\langle {\bf S}_{2} \rangle}{2m_2^2} + \frac{\langle {\bf S}_{1} \rangle + \langle {\bf S}_{2} \rangle}{m_1 m_2}\right) + \nonumber \\
&\quad +  \sum_{i,j=1}^3 \bigg[  \frac{1}{m_1m_2}\langle {{\bf S}}_{1} \rangle_i \langle {{\bf S}}_{2}\rangle_j + \nonumber \\
&\quad -  \frac{1}{2m_1^2} ({\bm \epsilon}_{1}^*)_i ({\bm \epsilon}_{1})_j  - (1 \to 2) \bigg] I_{ij}(r) \bigg\}.
\end{align}

%%%%%%%

\subsection{Partial conclusions}

\indent

It is interesting to notice the similarities between the bosonic cases treated. At the lowest order in $|{\bf p}|/m$ the potential energy is basically determined by $I_0(r)$. At this level the spin has no influence and the structure of the interaction is solely determined by the nature of the mediator, encoded in $a({\bf q})$. In this case one cannot use the potential energies to assess the nature of the sources, as these are indistinguishable. The monopole-monopole sector of the potentials for spin-0 and spin-1 sources is therefore identical.

For the static limit with spin-1/2 sources there is an extra term containing $I_1(r)$ (cf. eq.\eqref{Energia_fermion_2}). If we take a step back and redo the calculations, but starting from a truly static spinor, we would re-obtain the classic Yukawa potential $\sim I_0(r)$. This shows that this apparent discontinuity in the static approximation is actually an artifact of that very limit and is a particularity of spin-1/2 sources. In this sense, the potential energies of the three cases are equally proportional to the monopole-like $I_0(r)$ in the strict static limit (${\bf p} = 0$). This shows that, up to monopole-monopole terms, the interparticle energies $E^{(\rm s = 0)}(r)$, $E^{(\rm s = 1/2)}(r)$ and $E^{(\rm s = 1)}(r)$ are indistinguishable, what should not come as a surprise.

For non-static sources the situation is richer. At $\mathcal{O}(|{\bf p}|^2/m^2)$ we find a large interplay between momentum and spin. Familiar effects (e.g. spin-orbit coupling) appear and angular dependences become the rule. For well-known and/or carefully experimentally controlled (non-static) spin-polarized sources it is possible, then, to investigate the nature of the underlying interaction by direct inspection of the ensuing interparticle potential energy.

In the next sections, we apply the results above to specific interactions. First, however, we work out the familiar case of Maxwell electrodynamics. From now on we define $\alpha = e_1 e_2/4\pi$.

%%%%%%%%%%%%%%%%%%%%%%%%%%%%%%%

\section{Maxwell electrodynamics} \label{sec3}

\indent

Once we are interested in non-standard electrodynamics, it is important to establish common grounds for comparison. We start with the classical Maxwell Lagrangian, given by
\begin{eqnarray}
\L = -\frac{1}{4} F_{\mu\nu}F^{\mu\nu} -\frac{1}{2\lambda}(\pt_\mu A^\mu)^2, \label{L_max}
\end{eqnarray}
where $F_{\mu\nu} = \pt_\mu A_\nu - \pt_\nu A_\mu$ is the Abelian gauge-invariant field-strength tensor and the $\lambda$-dependent term fixes the gauge. This Lagrangian may be written as a quadratic form and its kernel, the wave operator, may be inverted to give the momentum-space propagator:
\begin{eqnarray}
\langle A_\mu A_\nu \rangle = - \frac{i}{q^2} \left[ \eta_{\mu\nu} + (\lambda - 1) \frac{q_\mu q_\nu}{q^2} \right].   \label{prop_max}
\end{eqnarray}

Comparing eq.\eqref{prop_max} with  eq.\eqref{prop_generic} we find that 
$a(\textbf{q}) = 1/\textbf{q}^2$. In this case, the relevant integrals can be easily evaluated from eqs.\eqref{I_0}-\eqref{I_ij} with $\xi = 0$ and the interparticle potential for scalar sources reads 
\begin{equation}
E^{\rm (s=0)}_{\rm Max}(r) = \frac{\alpha}{r} \left( 1  + \frac{\textbf{p}^2}{m_1 m_2} \right),
\end{equation}
which is compatible with the results from Refs.~\cite{Holstein1,Gupta}, apart from contact terms (cf. Section~\ref{sec_s0_cur}).

Similarly, for spin-1/2 sources we find
\begin{align}\label{Energia_fermion_maxwell}
& E^{\rm (s=1/2)}_{\rm Max}(r)  =   \frac{\alpha}{r} \,\bigg\{  1 + \frac{ \textbf{p}^2}{m_1 m_2}  + \nonumber \\ 
& -  \frac{1}{r^2} \textbf{L} \!\cdot\! \left( \frac{1}{2m_1^2} \langle \textbf{S}_1 \rangle \!+\! \frac{1}{2m_2^2} \langle \textbf{S}_2 \rangle \!+\! \frac{ \langle \textbf{S}_1 \rangle \!+\! \langle \textbf{S}_2 \rangle }{m_1 m_2} \right) + \nn \\  
& +  \frac{1}{m_1m_2 r^2} \left[ \langle {{\bf S}}_{1} \rangle \!\cdot\! \langle {{\bf S}}_{2}\rangle - 3 \left( \hat{ {\bf r} } \!\cdot\! \langle {{\bf S}}_{1} \rangle \right) \left( \hat{ {\bf r} } \!\cdot \!\langle {{\bf S}}_{2} \rangle \right) \right] \bigg\} + \nonumber \\ 
& -  4\pi\alpha \left[ \frac{2 \langle {{\bf S}}_{1} \rangle \cdot \langle {{\bf S}}_{2}\rangle }{3 m_1m_2}  \!+\!  \frac{1}{8\,m_1^2} \!+\! \frac{1}{8\,m_2^2}  \right] \delta^3(\textbf{r})
\end{align}
whereas, for spin-1 sources, we have
\begin{align}\label{Energia_spin_1_maxwell}
& E^{\rm (s=1)}_{\rm Max}(r)  =  \frac{\alpha}{r}  \bigg\{  1 + \frac{ \textbf{p}^2}{m_1 m_2} - \left( \frac{1}{2m_1^2} + \frac{1}{2m_2^2}\right) \frac{1}{r^2} +\nn\\
& -  \frac{1}{r^2} \textbf{L} \cdot \left(  \frac{1}{2 m_1^2} \langle \textbf{S}_1 \rangle   \!+\!  \frac{1}{2 m_2^2} \langle \textbf{S}_2 \rangle  \!+\!  \frac{ \langle \textbf{S}_1 \rangle  \!+\!  \langle \textbf{S}_2 \rangle }{m_1 m_2} \right) +\nn\\
& +  \frac{1}{m_1m_2 r^2} \left[  \langle \textbf{S}_1 \rangle \cdot \langle \textbf{S}_2 \rangle \!-\! 3 \left( \hat{ {\bf r} }\cdot \langle \textbf{S}_1 \rangle \right) \left( \hat{ {\bf r} }\cdot \langle \textbf{S}_2 \rangle \right) \right] +\nn\\
& +  \frac{3}{2 \,r^2} \left[ \frac{1}{m_1^2} ({\bm \epsilon}^*_1 \cdot \hat{ {\bf r}} ) ({\bm \epsilon}_1 \cdot \hat{ {\bf r}} )  \!+\!  (1\to 2) \right] \bigg\} +  \nn\\
& -  4\pi\alpha\left[ \frac{2}{3 m_1m_2} \langle {{\bf S}}_{1} \rangle \cdot \langle {{\bf S}}_{2}\rangle  \!+\!  \frac{1}{6\,m_1^2}  \!+\!  \frac{1}{6\,m_2^2}  \right] \delta^3(\textbf{r}). 
\end{align}

Overall, we see that some universalities between spin-1/2 and spin-1 sources are present, specially in the dipole-dipole (spin-spin) and momentum-dependent sectors of the respective interactions. As expected, the dominating contribution comes from the monopole-monopole term, which is the usual Coulomb one. We notice the appearance of contact terms, specifically the spin-spin one, which plays a role in multi-electron systems~\cite{Drake}.

From the form of eqs.\eqref{Energia_fermion_2} and \eqref{Energia_boson_2}, specially the respective second lines, we would expect extra spin- and momentum-independent contributions to $E^{\rm (s=1/2)}_{\rm Max}(r)$, but not to $E^{\rm (s=1)}_{\rm Max}(r)$. Contrary to our expectations, in eq.\eqref{Energia_spin_1_maxwell} there is another such term besides the Coulomb one. It arises as a by-product of the contractions of the polarization vectors -- they satisfy ${\bm \epsilon}_{i}^{\ast} \cdot{\bm \epsilon}_{i} = 1$ with $i=1,2$ -- and the Kr\"onecker deltas in $I_{ij}(r)$ (cf. eq.\eqref{I_ij}). Apart from contact terms, our results match those from Ref.~\cite{Holstein1}.

%%%%%%%%%%%%%%%%%%%%%%%%%%%%%%%

\section{Podolsky-Lee-Wick electrodynamics} \label{sec4}

\indent

Recently, interest in the Podolsky-Lee-Wick (PLW) electrodynamics has been renewed as Grinstein, O’Connell, and Wise extended the usual PLW model to a non-Abelian scenario and applied it to the Standard Model in an attempt to solve the hierarchy (Higgs mass) problem~\cite{Grinstein}-\cite{Alvarez}. Here, however, we shall focus on the simpler Abelian PLW electrodynamics.

More specifically, we wish to examine the role played by the (matter) sources interacting through the PLW fields. This topic has been the subject of many studies regarding, e.g.,
the point-like self energy of the electron \cite{Nogueira0}, monopoles \cite{MarioJr} and charged stationary branes and Dirac strings \cite{Nogueira}.

In this section we investigate the NR potential energy between particles interacting via the Abelian Podolsky-Lee-Wick (PLW) electrodynamics. Its dynamics is governed by the Lagrangian \cite{LW1,LW2}
\begin{equation}
\mathcal{L}_{\rm PLW} = -\frac{1}{4} F_{\mu\nu} \left( 1 + \frac{\Box}{M^2}  \right) F^{\mu\nu} + \frac{1}{2\lambda}(\partial_{\mu} A^{\mu}),
\end{equation}
so that one can immediately compute the Feynman propagator, which reads
\begin{equation}
D_{\mu\nu} = \frac{i M^2}{q^2 (q^2 - M^2)}\bigg\{ \eta_{\mu\nu} - \frac{q_\mu q_\nu}{k^2} \bigg[ 1 + \lambda \bigg(\frac{q^2}{M^2} - 1 \bigg) \bigg] \bigg\}.
\end{equation}

By inspecting the poles when contracted with external conserved currents, we conclude that the spectrum of the PLW electrodynamics consists of a massless unitary particle and of a non-unitary (ghost) particle with mass $M$~\cite{Pod1}-\cite{LW2}. Despite of its presence, the PLW model is still considered a good effective theory -- in particular, the PLW QED is naturally finite in the UV sector.

Usual Maxwell electrodynamics is recovered in the limit $M \to \infty$, which is physically sensible, as the ultra-heavy mode is not excited and plays no role in physical processes. Experimentally, one may find lower limits on $M$ by examining corrections to the $g$-factor of the electron, whereby it is found that $M > 40 \, {\rm GeV}$~\cite{Accioly3}. This shows that, if the PLW electrodynamics is realized in nature, it can only be significantly different from Maxwell's theory at distances $\ell_{\rm M} \sim 1/M \sim 10^{-16} \, {\rm cm}$ -- hundred times smaller than the classical radius of the electron.

Bearing in mind that conserved external currents are orthogonal to $k_\mu$, we promptly obtain the reduced propagator $\mathcal{P}_{\mu\nu}(k)$ as being
\begin{equation}
\mathcal{P}_{\mu\nu}(q) = -i\bigg(\frac{1}{q^2} - \frac{1}{q^2 - M^2}\bigg) \eta_{\mu\nu}, \label{p_PLW}
\end{equation}
thus concluding that $a(q) = -\frac{1}{q^2} + \frac{1}{q^2-M^2}$. The ghost nature of the massive particle is clear from the ``wrong'' sign of the massive piece in eq.\eqref{p_PLW}. As mentioned before, we are working in the limit of elastic scattering, where $q^0 = 0$, so that $a(\textbf{q})$ reduces to
\begin{equation}
a(\textbf{q}) = \frac{1}{\textbf{q}^{\,2}} -\frac{1}{\textbf{q}^{\,2} + M^2},\label{a_q}
\end{equation}
and we are now ready to specialize the discussion from Section~\ref{sec2} to the PLW electrodynamics.

%so that, using the results presented in Appendix \ref{integrals}, we obtain the following integrals:
%\begin{eqnarray} 
%I_0(r) & = & \frac{1}{4 \pi r}( 1 - e^{-M r})  \label{int_1_LW} \\
%I_1(r) & = & \frac{M^2}{4 \pi r} e^{-M r} \label{int_2_LW} \\
%\textbf{I}(r) & = & \frac{i}{4\pi r^2}\bigg[ 1 - (1 + M r) e^{-Mr}  \bigg] \hat{{\bf r}} \label{int_3_LW} \\
%I_{ij}(r) & = & \frac{1}{4\pi r^3} \bigg\{ \delta_{ij} - \delta_{ij}(1-M r)r^{-Mr} +  \nonumber \\
%&+& \frac{3x_i x_j}{r^2} \bigg[ 1 + \bigg(1 + M r + \frac{M^2 r^2}{3} \bigg) e^{-Mr}   \bigg] \bigg\}. \label{int_4_LW}
%\end{eqnarray}

%%%%%%%%%%%%%%%%%%%%%%%%%%%%

\subsection{Results}

\indent

Having set the basis of the model, we are ready to consider the simplest case, where the currents are described by (charged) spin-0 particles. Using eq.\eqref{energy_scalar} and eq.\eqref{I_0} with eq.\eqref{a_q}, we find 
\begin{eqnarray}
E^{\rm (s=0)}_{\rm PLW}(r) = \frac{\alpha}{r} \left( 1  + \frac{\textbf{p}^2}{m_1 m_2} \right)\left(1 - e^{-M r}\right), \label{en_scalar}
\end{eqnarray}
which is the leading-order correction.

Returning to the static case, we recover the well-known monopole-monopole, $M$-dependent, PLW interaction $\sim \left( 1 - e^{-Mr} \right)/r$. Also, in the limit where $M \to \infty$, the second term in eq.\eqref{en_scalar} is suppressed by the exponential and we re-obtain the Coulomb potential, as expected.

%%%%%%%%%%%%%%%%%%%%%%%%%%

We may now proceed to the more interesting cases of spin-1/2 and spin-1 sources. For the former, we find that the potential energy reads
\begin{align}\label{Energia_fermion_PLW}
&E^{\rm (s=1/2)}_{\rm PLW}(r)=\frac{\alpha}{r} \bigg\{ \left( 1 + \frac{ \textbf{p}^2}{m_1 m_2}  \right) \left(1 - e^{-M r}\right) +\nn \\ 
& -  \frac{M^2}{8} \left( \frac{1}{m_1^2} + \frac{1}{m_2^2} \right) e^{-Mr}  + \nn \\
& -  \frac{f_1(r) }{r^2} \,\textbf{L} \!\cdot\! \left[ \frac{1}{2m_1^2} \langle \textbf{S}_1 \rangle \!+\! \frac{1}{2m_2^2} \langle \textbf{S}_2 \rangle \!+\! \frac{ \langle \textbf{S}_1 \rangle \!+\! \langle \textbf{S}_2 \rangle }{m_1 m_2} \right] + \nn \\
& +  \frac{1}{m_1m_2 r^2} \bigg[ f_2(r)  \langle \textbf{S}_1 \rangle \cdot \langle \textbf{S}_2 \rangle + \nn \\
& -  3f_3(r) \left( \hat{ {\bf r} }\cdot \langle \textbf{S}_1 \rangle \right) \left( \hat{ {\bf r} } \cdot \langle \textbf{S}_2 \rangle \right) \bigg] \bigg\},
\end{align}
where we defined $f_1(r) = 1 - (1 + Mr) e^{-M r}$, $f_2(r) = 1 - (1 + M r + M^2 r^2)e^{-M r}$ and $f_3(r) = 1 - (1 + M r + \frac{M^2 r^2}{3})e^{-M r}$ for the sake of convenience.

Using the definitions above we may write the potential energy for spin-1 sources in the PLW electrodynamics as
\begin{align}\label{Energia_boson_PLW}
&E^{\rm (s=1)}_{\rm PLW}(r) = \frac{\alpha}{r} \bigg\{ \left( 1 + \frac{ \textbf{p}^2}{m_1 m_2}  \right) \left(1 - e^{-M r}\right) +\nn \\ 
& -  \frac{f_1(r)}{2r^2} \left( \frac{1}{m_1^2} + \frac{1}{m_2^2} \right) e^{-Mr}  + \nn \\
& -  \frac{f_1(r) }{r^2} \,\textbf{L} \!\cdot\! \left[ \frac{1}{2m_1^2} \langle \textbf{S}_1 \rangle \!+\! \frac{1}{2m_2^2} \langle \textbf{S}_2 \rangle \!+\! \frac{ \langle \textbf{S}_1 \rangle \!+\! \langle \textbf{S}_2 \rangle }{m_1 m_2} \right] + \nn \\
& +  \frac{1}{m_1m_2 r^2} \bigg[ f_2(r)  \langle \textbf{S}_1 \rangle \cdot \langle \textbf{S}_2 \rangle + \nn \\
& -  3f_3(r) \left( \hat{ {\bf r} }\cdot \langle \textbf{S}_1 \rangle \right) \left( \hat{ {\bf r} } \cdot \langle \textbf{S}_2 \rangle \right) \bigg] +\nn \\ 
& +  \frac{3f_3(r)}{2r^2} \left[ \frac{1}{m_1^2} ({\bm \epsilon}^*_1 \cdot \hat{ {\bf r} } ) ({\bm \epsilon}_1 \cdot \hat{ {\bf r} } ) + (1 \to 2) \right] \bigg\}.
\end{align}

A remarkable consequence of the relative sign in eq.\eqref{a_q} is that the integrals in Appendix~\ref{integrals} present no contact terms $\sim \delta^3 ({\bf r})$. This is only possible due to the particular structure of the PLW propagator, eq.\eqref{a_q}, which allows a natural cancellation. This is a very distinctive feature of the interaction energies in the PLW electrodynamics, when compared to the standard Maxwell one.

%%%%%%%%%%%%%%%%%%%%%%%%%%%%%%%

\section{Kinetically mixed photon-hidden-photon electrodynamics} \label{sec5}

\indent

Let us now apply the formalism developed in Section~\ref{sec2} to the case where the usual Maxwell electrodynamics is extended by the inclusion of an extra Abelian boson without direct interaction with the matter sector. This novel boson, $B_\mu$, interacts with the standard photon only through a kinetic mixing. The gauge Lagrangian is then 
\begin{eqnarray}
\L = -\frac{1}{4}F_{\mu\nu}^2 - \frac{1}{4}B_{\mu\nu}^2 + \frac{\chi}{2}B_{\mu\nu}F^{\mu\nu} + \frac{m_{\gamma\prime}^2}{2}B_{\mu}^2 , \label{L_hd}
\end{eqnarray}
where $B_{\mu\nu} = \pt_\mu B_\nu - \pt_\nu B_\mu$ is the field-strength tensor of the hidden photon ($\gamma^\prime$), whose mass is $m_{\gamma^\prime}$.

The hidden photon is completely decoupled from the visible sector except for the kinetic mixing term, $\frac{\chi}{2}B_{\mu\nu}F^{\mu\nu}$. Here we shall take this term as a true interaction vertex, 
\begin{equation}
V^{\mu\nu}_{\gamma - \gamma^\prime} = i \chi \left( \eta^{\mu\nu}\, q^2 - q^\mu q^\nu \right),
\end{equation}
so that, from eq.\eqref{L_hd} we may read the propagator of the hidden photon:
\begin{equation}
\langle B_\mu B_\nu \rangle = - \frac{i}{q^2-m^2_{\gamma^\prime}} \left( \eta_{\mu\nu} - \frac{q_\mu q_\nu}{m^2_{\gamma^\prime}} \right),
\end{equation}
while the propagator for the photon is given in eq.\eqref{prop_max}.

Given that the matter sources are not charged under $U(1)_B$ they can only feel the influence of the hidden photon by means of small corrections to the usual electromagnetic interaction. According to Ref. \cite{JJ}, the parameter $\chi$ is constrained to $ 10^{-12} \lesssim \chi \lesssim 10^{-3}$, so we may write the effective photon propagator as $\langle A_\mu A_\nu \rangle_{\rm eff} = \langle A_\mu A_\alpha \rangle V^{\alpha\rho}_{\gamma - \gamma^\prime} \langle B_\rho B_\lambda \rangle V^{\lambda\sigma}_{\gamma - \gamma^\prime} \langle A_\sigma A_\nu \rangle + \cdots$, which can be recast as
\begin{eqnarray}
\langle A_\mu A_\nu \rangle_{\rm eff} = -i\left( \frac{1}{q^2} + \frac{\chi^2}{q^2-m_{\gamma^\prime}^2}\right) \eta_{\mu\nu} + iX_{\mu\nu}(\lambda), \label{eff_prop}
\end{eqnarray} 
with the last (gauge-dependent) piece $\sim X_{\mu\nu}(\lambda)$ vanishing upon contraction with conserved currents.

Comparing eq.\eqref{eff_prop} with eq.\eqref{prop_generic}, we find ($q^0 = 0$)
\begin{equation}
a(\textbf{q}) = \frac{1}{\textbf{q}^2} + \frac{\chi^2}{\textbf{q}^2+m_{\gamma^\prime}^2}, \label{a_q_hd}
\end{equation}
which is similar to eq.\eqref{a_q} for the PLW electrodynamics, but it has an important difference: there is no relative sign, so we cannot expect a cancellation of the contact terms as in Section~\ref{sec4}.

%%%%%%%%%%%%%%%%%%%%%%%%%%%%%

\subsection{Results}

\indent

Evaluating the integrals in the Appendix~\ref{integrals} with eq.\eqref{a_q_hd} we find that the interaction energy between scalars is given by
\begin{equation}
E^{\rm (s=0)}_{\gamma - \gamma^\prime}(r) = \frac{\alpha}{r} \left( 1  + \frac{\textbf{p}^2}{m_1 m_2} \right)\left(1 + \chi^2 e^{- m_{\gamma^\prime} r}\right),
\end{equation}
whereas, for spin-1/2 sources, we have
\begin{align}\label{Energia_fermion_hd}
&E^{\rm (s=1/2)}_{\rm \gamma - \gamma^\prime}(r)  =  \frac{\alpha}{r} \bigg\{ \left( 1 + \frac{ \textbf{p}^2}{m_1 m_2}  \right) \left(1 + \chi^2 e^{- m_{\gamma^\prime} r}\right) +\nn \\ 
& +  \frac{ \chi^2 m_{\gamma^\prime}^2 }{8} \left( \frac{1}{m_1^2} + \frac{ 1}{m_2^2} \right)  e^{-m_{\gamma^\prime} r}  +\nn \\
& -  \frac{g_1(r) }{r^2} \,\textbf{L} \!\cdot\! \left[ \frac{1}{2m_1^2} \langle \textbf{S}_1 \rangle \!+\! \frac{1}{2m_2^2} \langle \textbf{S}_2 \rangle \!+\! \frac{ \langle \textbf{S}_1 \rangle \!+\! \langle \textbf{S}_2 \rangle }{m_1 m_2} \right] + \nn \\
& +  \frac{1}{m_1m_2 r^2} \bigg[ g_2(r)  \langle \textbf{S}_1 \rangle \cdot \langle \textbf{S}_2 \rangle + \nn \\
& -  3g_3(r) \left( \hat{ {\bf r} }\cdot \langle \textbf{S}_1 \rangle \right) \left( \hat{ {\bf r} } \cdot \langle \textbf{S}_2 \rangle \right) \bigg] \bigg\} + \nn\\
& -  4\pi\tilde{\alpha} \left[ \frac{2}{3 m_1m_2} \langle {{\bf S}}_{1} \rangle \!\cdot\! \langle {{\bf S}}_{2}\rangle \!+\!  \frac{1}{8\,m_1^2} \!+\! \frac{1}{8\,m_2^2}  \right] \delta^3(\textbf{r}), 
\end{align}
where we used $\tilde{\alpha} = \alpha (1+\chi^2)$ and defined the functions $g_1(r) = 1 + \chi^2(1 + m_{\gamma^\prime}r) e^{-m_{\gamma^\prime} r}$, $g_2(r) = 1 + \chi^2 (1 + m_{\gamma^\prime} r + m^2_{\gamma^\prime} r^2 )e^{-m_{\gamma^\prime} r}$ and $g_3(r) = 1 + \chi^2 \left(1 + m_{\gamma^\prime}r + \frac{ m_{\gamma^\prime}^2 r^2}{3} \right)e^{-m_{\gamma^\prime} r}$.

Finally, the interaction energy between spin-1 sources in the hidden-photon electrodynamics is
\begin{align}\label{Energia_boson_hd}
&E^{\rm (s=1)}_{\rm \gamma - \gamma^\prime}(r) = \frac{\alpha}{r} \bigg\{ \left( 1 + \frac{ \textbf{p}^2}{m_1 m_2}  \right) \left(1 + \chi^2 e^{- m_{\gamma^\prime} r}\right) + \nn \\ 
& -  \frac{ g_1(r) }{2r^2} \left( \frac{1}{m_1^2} + \frac{ 1}{m_2^2} \right)  e^{-m_{\gamma^\prime} r}  +\nn \\
& -  \frac{g_1(r) }{r^2} \,\textbf{L} \!\cdot\! \left[ \frac{1}{2m_1^2} \langle \textbf{S}_1 \rangle \!+\! \frac{1}{2m_2^2} \langle \textbf{S}_2 \rangle \!+\! \frac{ \langle \textbf{S}_1 \rangle \!+\! \langle \textbf{S}_2 \rangle }{m_1 m_2} \right] + \nn \\
& +  \frac{1}{m_1m_2 r^2} \bigg[ g_2(r)  \langle \textbf{S}_1 \rangle \cdot \langle \textbf{S}_2 \rangle + \nn \\
& -  3g_3(r) \left( \hat{ {\bf r} }\cdot \langle \textbf{S}_1 \rangle \right) \left( \hat{ {\bf r} } \cdot \langle \textbf{S}_2 \rangle \right) \bigg] \nn \\
& +  \frac{3g_3(r)}{2r^2} \left[ \frac{1}{m_1^2} ({\bm \epsilon}^*_1 \cdot \hat{ {\bf r} } ) ({\bm \epsilon}_1 \cdot \hat{ {\bf r} } ) + (1 \to 2) \right] \bigg\} +\nn \\
& -  4\pi\tilde{\alpha} \left[ \frac{2}{3 m_1m_2} \langle {{\bf S}}_{1} \rangle \!\cdot\! \langle {{\bf S}}_{2}\rangle \!+\!  \frac{1}{6\,m_1^2} \!+\! \frac{1}{6\,m_2^2}  \right] \delta^3(\textbf{r}). 
\end{align}

The lowest-order contribution from eq.\eqref{Energia_fermion_hd} matches the interaction potential used in Ref.~\cite{JJ} to extract limits on the $\chi - m_{\gamma^\prime}$ parameter space through spectroscopy measurements.

%Here we highlight another possibility to study these modified electrodynamics. One may find the effective action for the electromagnetic field by integrating out the $B_{\mu}$-field in eq.\eqref{L_hd}. This was carried out in Ref.~\cite{Gaete_Schmidt} for static spin-1/2 sources and the spin-independent (static) potential obtained coincides with our first correction up to $\mathcal{O}(\chi^2)$.

%%%%%%%%%%%%%%%%%%%%%%%%%%%%

\section{Concluding remarks} \label{sec6}

\indent

We have systematically analyzed the interparticle potential energy between sources of different spins in the context of modified electrodynamics. For concreteness, we treated the well-known case of Maxwell electrodynamics and of two extensions, namely, that of Podolsky-Lee-Wick and that with an extra Abelian dark gauge boson (hidden photon).

We worked with NR amplitudes up to $\mathcal{O}(|{\bf p}|^2/m^2)$, whereby a broad variety of spin- and velocity-dependent terms arise, including well-known effects such as spin-orbit couplings, as well as more exotic ones involving the polarizations. For the modified electrodynamics, the ensuing interparticle potentials include Yukawa-like terms with a typical interaction range $\ell \sim 1/\xi$, where $\xi = M$ or $m_{\gamma^\prime}$ is the mass of the mediator (PLW and hidden-photon, respectively).

%As mentioned in Section~\ref{sec_s0}, the PLW electrodynamics may be subjected to phenomenological tests: its mass scale must be above the GeV scale \cite{Accioly3}, so let us consider the case where the mass parameter $M$ is large, but finite. For definiteness we will assume that particles 1 and 2 are electrons ($m_1 = m_2 \equiv m_e = 0.5 \, {\rm MeV}$ and $e_1 = e_2 \equiv -e$). If we take the fully static limit, we end up with $J_{1 (2)}^0 = e$ and ${\bf J}_{1(2)} \approx 0$, so that
%\begin{equation}
%E(r) = \frac{e^2}{4 \pi r} (1 - e^{-M r}), \label{static}
%\end{equation} 
%but this is not what is found if we take ${\bf p} \to 0$ and ${\bf q} \to 0$ in eqs.\eqref{E_0} and \eqref{E_s_vel}; we neglect spin effects for simplicity. A similar result is found for the spin-0 case.

%In Ref.~\cite{IF} the authors have obtained a multipole expansion for the PLW electrodynamics with static localized sources. In their treatment, as is usual in classical electromagnetism, electric charges and closed-loop circuits with steady currents (physical dipoles) are treated separately. Here, as we give a microscopic treatment of point charges (with and without spin), the interaction energy contain mixed contributions from their results for point charges and current loops. 

%COMPLETE DISCUSSION

Another possibility to study the modified photon-hidden-photon electrodynamics is by finding the effective action for the electromagnetic field by integrating out the $B_{\mu}$-field in eq.\eqref{L_hd}. This was carried out in Ref.~\cite{Gaete_Schmidt}  and the spin-independent (static) potential obtained coincides with our first correction up to $\mathcal{O}(\chi^2)$.

In the two applications we presented there are spin- and momentum-independent terms in the potential energies for both spin-1/2 and spin-1 sources. These terms are divided in two classes: $\sim \xi^2/r$, with $\xi = M$ or $m_{\gamma^\prime}$, for spin-1/2 (cf. eqs.\eqref{Energia_fermion_PLW} and \eqref{Energia_fermion_hd}) and $\sim 1/r^3$ for spin-1 sources (cf. eqs.\eqref{Energia_boson_PLW}  and \eqref{Energia_boson_hd}), both accompanied by a factor of $m_1^{-2} + m_2^{-2}$. The origins of these terms is quite different. For spin-1/2 sources it is easy to see that these factors come from the second term in eq.\eqref{Energia_fermion_2}, since $I_1 (r) \sim \xi^2/r$ (cf. eq.\eqref{I_1}). The origin of such terms for spin-1 sources was already indicated in the end of Section~\ref{sec3}: it arises as part of the contraction $({\bm \epsilon}_{1}^*)_i ({\bm \epsilon}_{1})_j I_{ij}(r) + (1 \to 2)$ in eq.\eqref{Energia_boson_2}, since $I_{ij}(r) \sim \delta_{ij}$ (cf. eq.\eqref{I_ij}) and the polarization 3-vectors satisfy $({\bm \epsilon}_{1,2}^*)_k ({\bm \epsilon}_{1,2})_k = 1$.

As indicated in the end of Section~\ref{sec2}, the contact terms arise only when the NR amplitude is expressed as a series expansion in $|{\bf p}|/m$. The same is also true not only for the momentum-dependent terms, but also for the ones containing information on the spin (or polarization) of the sources. For fermions exchanging the usual photon, this is a direct consequence of the fact that the spin operator only appears as ${\bm \sigma}\cdot{\bf p}/m$ (cf. eq.\eqref{spinor_basic}). If we start with truly static sources, the Coulomb potential $\sim \alpha/r$ is dully recovered. Similar conclusions apply for the modified electrodynamics discussed above.

Although not explicit, an analogous situation is found for the electromagnetic interaction of spin-1 sources, as can be seen by the profile $\sim 1/r^3$ of the third term in eq.\eqref{Energia_spin_1_maxwell}. Since $1/r^3 \sim {\bf q}^3$ in Fourier space, we see that this term actually stems from $I_{ij}(r)$, eq.\eqref{I_ij}, with $a({\bf q}) = 1/{\bf q}^2$, so this too is a reminiscent of the series expansion in powers of $|{\bf p}|/m$. As in the spin-1/2 case, starting directly with ${\bf p} = 0$, the only term that survives in eq.\eqref{Energia_spin_1_maxwell} is the first. This shows that, irrespective of their spin, static sources interact via the Coulomb potential $\alpha/r$.

In summary, the analysis presented in Section~\ref{sec2} and implemented in the following sections is general and may be applied to any modification of standard electrodynamics which keeps Lorentz invariance; the main requirement is that the propagator may be decomposed as in eq.\eqref{prop_generic}. Moreover, the results given above may be used to experimentally search for new mediators, as they would induce exotic spin- and velocity-dependent forces that could be detected in experiments involving e.g. torsion pendula~\cite{Adelberger}, rare earth iron magnets~\cite{Long}, geomagnetic electrons~\cite{Hunter} or magnetometers~\cite{Savukov}.

%Interestingly enough, if we take $|{\bf p}|/m \to 0$, specially in eq.\eqref{E_s_vel}, we find an extra contribution, so that the interparticle potential becomes
%\begin{equation}
%E(r) = \frac{e^2}{4\pi r}\left[ 1 - \left( 1 +  \frac{M^2}{2m^2} \right)e^{-Mr}  \right] + \mathcal{O}(|{\bf p}|/m) \label{approx}
%\end{equation}
%and we need to conciliate this with eq.\eqref{static} -- similar results are found in e.g. ref.\cite{helayel2}. This discontinuity in the static case has its origins in the amplitude, c.f.

%solution lies in the hierarchy among the different masses and energy scales involved, as well as on the approximations used. 

%The NR approximation is based solely on properties of the sources, i.e., the relation between $|{\bf p}|$ (or $|{\bf q}|$) and $m_e$ -- it does not take the other mass scale, $M$, explicitly into account. Eq.\eqref{approx} is obtained exactly within this premise, that is, by taking $|{\bf p}|/m \to 0$ only in the final result, the potential energy $E(r)$, whereas eq.\eqref{static} incorporates the NR approximation as early as in the current level.

%%%%%%

\section*{Acknowledgements}

\indent 

The authors are grateful to J. A. Helayël-Neto for
reading the manuscript. This work was funded by the
Brazilian funding agencies CNPq and FAPERJ, and the German Service for Academic Exchange (DAAD). P.C.M. would also like
to thank the Institut für theoretische Physik (U. Heidelberg) for the hospitality.

%%%%%%%%%%%%%%%%%%%%%%%%%%%%%%%%%%%%%%%%%%%%

\appendix

\section{Useful integrals \label{integrals}}

\indent

Throughout this paper we have used three classes of integrals, namely
\begin{eqnarray}
I_n(r) & = &  \int \frac{d^3 \textbf{q}}{(2\pi)^3}  \, (\textbf{q}^{\,2})^n \, a(\textbf{q})\,e^{i \textbf{q}\cdot\textbf{r}} \\
I_{ij}(r) & = & \int \frac{d^3 \textbf{q}}{(2\pi)^3} \, {\bf q}_i{\bf q}_j \, a(\textbf{q})\,e^{i \textbf{q}\cdot\textbf{r}} , \,\, (i,j = 1,2,3)
\end{eqnarray}

%and
%\begin{equation}
%\textbf{I}(r) = \int \frac{d^3 \textbf{q}}{(2\pi)^3} \, \textbf{q} \, a(\textbf{q})\,e^{i \textbf{q}\cdot\textbf{r}} .
%\end{equation}

%Since 3-dimensional rotational invariance requires that $a(\textbf{q})$ depends only on $|\textbf{q}|$, one can use spherical coordinates and integrate on the angular variables in order to obtain \cite{Accioly2}
%\begin{equation}\label{integral_n}
%I_n(r) =  \frac{1}{2\pi^2 r} \int_0^\infty dx \, x^{2n+1} a(x) \sin(xr),
%\end{equation}
%where we denote $x \equiv |\textbf{q}|$. 

It is not difficult to see that $I_{n+1}(r) = -\nabla^2 I_n(r)$ and $I_{ij}(r) = - \partial_i \partial_j I_0(r)$. Consequently, once we specify $a(\textbf{q})$, the only integral we have to worry about is $I_0(r)$.

%and
%\begin{equation}
%\textbf{I}(r) = -i\, \vec{\nabla} I_0(r) .
%\end{equation} 

%According to eq.\eqref{integral_n}, $I_0(r)$ can be written as 
%\begin{equation}\label{integral_0}
%I_0(r) =  \frac{1}{2\pi^2} \frac{1}{r} \int_0^\infty dx \, x \, a(x) \sin(xr).
%\end{equation}

In this paper we came across situations where $a(x) = \frac{1}{x^2 + \xi^2}$, with 
$\xi$ a real constant. Therefore, it will be useful to compute the integrals above in this specific case. Below we quote the results used in the main text (the limit $\xi \to 0$ may be taken):
\begin{eqnarray}
I_0(r) = \frac{1}{4 \pi r} e^{-\xi r} \label{I_0}
\end{eqnarray}

\begin{eqnarray}
I'_0(r) = - \frac{1}{4\pi r^2}\left( 1 + \xi r \right)e^{-\xi r} \label{I_0'}
\end{eqnarray}

\begin{eqnarray}
I_1 (r) =  \delta^3({\bf r}) - \frac{\xi^2}{4 \pi r} e^{-\xi r} \label{I_1}
\end{eqnarray}

\begin{align}
I_{ij}(r) = \frac{1}{3}\delta_{ij} \delta^3(\textbf{r})  & + \frac{1}{4\pi r^3} \bigg[ (1 + \xi r) \delta_{ij} + \label{I_ij} \\ &- (3 + 3 \xi r + \xi^2 r^2) \frac{x_i x_j}{r^2} \bigg] e^{-\xi r},\nn  
\end{align}

%\begin{eqnarray}
%\textbf{I}(r) =  \frac{i\,\alpha}{4 \pi r^2}(1 + r \xi) e^{-\xi r} \hat{{\bf r} }.
%\end{eqnarray}

The presence of the Dirac delta in eq.\eqref{I_ij} is justified once we recognize that $I_1 (r) = {\rm Tr}\left\{ I_{ij} (r) \right\}$ \cite{Ints}.

% \begin{eqnarray}
% I_0(r) = 
% \begin{cases}
% \frac{\alpha}{4 \pi r}, \quad& \textmd{if} \,\,\, \xi = 0 \\
% \frac{\alpha}{4 \pi r} e^{-\xi r}, \quad& \textmd{if} \,\,\, \xi \neq 0
% \end{cases}
% \end{eqnarray}

% \begin{eqnarray}
% I_1(r) = 
% \begin{cases}
% \alpha \, \delta({\bf r}), \quad& \textmd{if} \,\,\, \xi = 0 \\
% \alpha \, \left( \delta({\bf r}) -\frac{\xi^2}{4 \pi r} e^{-\xi r} \right), \quad& \textmd{if} \,\,\, \xi \neq 0
% \end{cases}
% \end{eqnarray}

% \begin{eqnarray}
% \textbf{I}(r) = 
% \begin{cases}
% \frac{i\,\alpha}{4 \pi r^2} \hat{{\bf r} }, \quad& \textmd{if} \,\,\, \xi = 0 \\
% \frac{i\,\alpha}{4 \pi r^2}(1 + r \xi) e^{-\xi r} \hat{{\bf r} }, \quad& \textmd{if} \,\,\, \xi \neq 0
% \end{cases}
% \end{eqnarray}

% \begin{eqnarray}
% I_{ij}(r) = 
% \begin{cases}
% \frac{\alpha}{3}\delta_{ij}\delta({\bf r}) + \frac{\alpha}{4\pi r^3} \left( \delta_{ij} - 3 \frac{x_i x_j}{r^2} \right), & \textmd{if} \,\, \xi = 0\\
% \frac{c}{4\pi r^3} \left( A \frac{x_i x_j}{r^2}  - B \delta_{ij} \right) e^{-\xi r}, &  \textmd{if} \,\, \xi \neq 0
% \end{cases}
% \end{eqnarray}
% where $A = 3 + 3 r \xi + r^2 \xi^2 $ and $B = 1 - r \xi$ and $i \neq j$ \cite{Ints}. We also note that $I_1 (r) = {\rm Tr}\big\{ I_{ij}(r) \big\}$.

%%%%%%%%%%%%%%%%%%%%%%%%%%%%%%%%%%%%%%%%%%%%%%%%%


\begin{thebibliography}{99}


%%%%%%% INTRO %%%%%%%%

%\bibitem{Jackson} J.D. Jackson, {\it Classical electrodynamics}, $3^{\rm rd}$ edition, Ed. Wiley.

\bibitem{Holstein1} B.R. Holstein, A. Ross, arXiv: hep-ph/0802.0715v1.

\bibitem{Holstein2} B.R. Holstein, arXiv: hep-ph/1609.00714.

\bibitem{Holstein3} B.R. Holstein, A. Ross, arXiv: hep-ph/0802.0716.

\bibitem{Nogueira0} F.A. Barone, G. Flores-Hidalgo, A.A. Nogueira, Phys. Rev. D {\bf 91}, 027701 (2015).

\bibitem{MarioJr} R. Turcati, M.J. Neves, Adv. High Energy Phys. {\bf 24}, 153953 (2014).

\bibitem{Nogueira} F.A. Barone, G. Flores-Hidalgo, A.A. Nogueira, Phys. Rev. D {\bf 88}, 105031 (2013).

%\bibitem{helayel2} P.C. Malta, L.P.R. Ospedal, K. Veiga and J.A. Helay\"{e}l-Neto, Adv. High Energy Phys. {\bf 2016}, 2531436 (2016).

\bibitem{Maggiore} M. Maggiore, \textit{A modern introduction to quantum field theory}, Oxford University Press (2005).

\bibitem{Pod1} B. Podolsky, Phys. Rev.{\bf 62}, 68 (1942).

\bibitem{Pod2} B. Podolsky, Phys. Rev. {\bf 65}, 228 (1944).

\bibitem{LW1} T. Lee, G. Wick, Nucl. Phys. B \textbf{9}, 209 (1969).

\bibitem{LW2} T. Lee, G. Wick, Phys. Rev. D \textbf{2}, 209 (1970).

\bibitem{Holdon} B. Holdom, Phys. Lett. B {\bf 166}, 196 (1986).

\bibitem{JJ} J. Jaeckel, S. Roy, Phys. Rev. D {\bf 82}, 125020 (2010).





%%%%%% SEC 2 %%%%%%%%%

\bibitem{epjc} F.A. Gomes Ferreira, P.C. Malta, L.P.R. Ospedal, J.A. Helayël-Neto, Eur. Phys. J. C {\bf 75}, 238 (2015). 

\bibitem{Accioly1} A. Accioly, J. Helay\"{e}l-Neto, F.E. Barone, F.A. Barone and P. Gaete, Phys. Rev. D \textbf{90}, 105029 (2014).

\bibitem{Gupta} S.N. Gupta, S.F. Radford, Phys. Rev. D {\bf 21}, 2213 (1980).

\bibitem{Drake} G.W.F. Drake (ed.), {\it Springer handbook of atomic, molecular and optical physics}, Springer Ed. (2006).

%%%%%% SEC 3 %%%%%%%%%



%%%%%% SEC 4 %%%%%%%%%

\bibitem{Grinstein} B. Grinstein, D. O'Connell, M.B. Wise, Phys. Rev. D {\bf 77}, 025012 (2008).

\bibitem{Rizzo} T.G. Rizzo, JHEP {\bf 06}, 070 (2007).

\bibitem{Alvarez} E. \`Alvarez, C. Schat, L. da Rold, A. Szynkman, JHEP {\bf 04}, 026 (2008).



%\bibitem{Zee} A. Zee, \textit{Quantum Field Theory in a Nutshell}, 2nd ed. (Princeton University Press, Princeton, NJ, 2010).

%\bibitem{Helayel1} F.A. Gomes Ferreira, P.C. Malta, L.P.R. Ospedal and J.A. Helay\"{e}l-Neto, Eur. Phys. J. C \textbf{75}, 238 (2015).

 
\bibitem{Jackiw} R. Jackiw, Phys. Rev. D \textbf{57}, 2635 (1998).

\bibitem{Ferrara} S. Ferrara, M. Porrati and V. Teledgi, Phys. Rev. D \textbf{46}, 3529 (1992).

\bibitem{Accioly2} A. Accioly, J. Helay\"el-Neto, G. Correa, G. Brito, J. de Almeida and W. Herdy, Phys. Rev. D \textbf{93}, 105042 (2016).

\bibitem{Accioly3} A. Accioly, E. Scatena, Mod. Phys. Lett. A {\bf 25}, 269-276 (2010). 

%\bibitem{IF} C.A. Bonin, B.M. Pimentel, P.H. Ortega, arXiv:hep-th/1608.00902v1.

\bibitem{Gaete_Schmidt} Patricio Gaete, Ivan Schmidt,
 Int. J. Mod. Phys. A {\bf 26}, 863 (2011).

\bibitem{Adelberger} W.A. Terrano, E.G. Adelberger, J.G.Lee, B.R. Heckel, Phys. Rev. Lett. {\bf 115}, 201801 (2015).

\bibitem{Long} T.M. Leslie, E. Weisman, R. Khatiwada, J.C. Long,
Phys. Rev. D {\bf 89}, 114022 (2014).

\bibitem{Hunter} L.R. Hunter, D.G. Ang, Phys. Rev. Lett. {\bf 112}, 091803 (2014).

\bibitem{Savukov} P.H. Chu, Y.J. Kim, I. Salukov, Phys. Rev. D {\bf 94}, 036002 (2016).



%%%%%%%%%%%% APENDICE %%%%%%%%%%%%%%%

\bibitem{Ints} G.S. Adkins, {\it Three-dimensional Fourier transforms, integrals of spherical Bessel functions, and novel delta function identities}, arXiv: math-ph/1302.1830v1.


\end{thebibliography}
\end{document}